\documentclass{PoS}

\title{Magnetic Reconnection, Cosmic Ray Acceleration, and Gamma-Ray emission around Black Holes and Relativistic Jets}

\ShortTitle{Magnetic Reconnection, Acceleration, and Gamma-Rays }

\author{\speaker{Elisabete M. de Gouveia Dal Pino}\\
       Instituto de Astronomia, Geof\'isica e Ci\^encias Atmosf\'ericas - Universidade de S\~{a}o Paulo (IAG/USP), S\~{a}o Paulo, Brazil\\
        E-mail: \email{dalpino@iag.usp.br}}
\author{Rafael Alves Batista \\
        IAG - Universidade de S\~{a}o Paulo, Brazil\\}

\author{Luis S. Kadowaki\\
        IAG - Universidade de S\~{a}o Paulo, Brazil\\}
\author{Grzergorz Kowal\\
        EACH - Universidade de S\~{a}o Paulo, and UniSul, Brazil\\}

\author{Tania Medina-Torrejon\\
        Instituto de Fisica\ - Universidade de S\~{a}o Paulo, Brazil\\}

\author{Juan Carlos Ramirez-Rodriguez\\
       IAG - Universidade de S\~{a}o Paulo, Brazil\\}
       
\abstract{Particle acceleration by magnetic reconnection is now recognized as an important process in magnetically dominated regions of galactic and extragalactic  black hole sources.  This process helps to solve current puzzles specially related to the origin of the very high energy flare emission  in  these sources.  In this review, we discuss  this acceleration mechanism and show recent analytical studies and multidimensional numerical SRMHD and GRMHD (special and general relativistic magnetohydrodynamical)  simulations with the injection of test particles, which  help us to understand  this process  both in  relativistic jets and coronal regions of these sources. The very high energy and neutrino emission resulting from the accelerated particles by reconnection is also discussed.}

\FullConference{International Conference on Black Holes as Cosmic Batteries: UHECRs and Multimessenger Astronomy - BHCB2018\\
		12-15 September, 2018\\
		Foz du Iguazu, Brasil}

\begin{document}

\section{Introduction}
%Black Holes (BH) and associated accretion and relativistic jet phenomena are usually evidenced   in galactic binary systems (or BHBs, or microquasars). Also, supermassive BHs (with   $\sim 10^6$ to $10^{10}$  solar masses) are believed to occur in the nuclei  of all classes of active galaxies (or AGNs, from the high luminous  blazars to the less luminous   radio or  seyfert galaxies). Besides, it is also argued that BHs may be the ultimately engines of gamma-ray-burts (GRBs).

%Multi-wavelength observations  indicate that the collimated relativistic jets produced near the central BH can be accelerated to large Lorentz factors,  up to several orders of magnitude in length scales. Their formation is still a matter of debate, but the usually  most accepted models rely on magnetic processes, like  magneto-centrifugal acceleration by helical magnetic fields arising from the accretion disk \cite{blandford_payne82}. Or otherwise  they can be powered by the BH spin transferred to the surrounding magnetic flow  \cite{Blandford_Znajek1977}. 
%Thus in any case the  prediction is  that the jets should be  born as magnetically  or  Poynting flux dominated flows. The fact that at  observable distances from the source (starting  around $1000 R_S$ or less, where $R_S$ is the Schwartzschield radius) these jets become kinetically dominated, indicate that they must somehow suffer an efficient  conversion (or dissipation) of magnetic into kinetic energy. We  argue below that magnetic reconnection may be a powerful mechanism operating in these jets to allow for this conversion. 

The detection of gamma-ray flares at TeV energies in AGN blazars, like PKS2155-304 \cite{ahar07}, and more recently, the simultaneous observation of an IceCube neutrino event  and a  gamma-ray
flare in  blazar TXS 0506+056 \cite{IceCube18}  are among the most puzzling  discoveries in high-energy astrophysics.  
%The very high variability in intensity and short duration of these flares are suggestive of explosive and very compact emission regions. 
%Such very high energy (VHE) flares have been also detected in other compact sources, like the Crab pulsar wind nebula  (Abdo et al. 2011; Tavani et al. 2011; Lyutikov et al. 2017).
%, and Cyg X1 and X3 galactic black hole X-ray binaries (Khiali et al. 2015 and references therein).

Blazars, which are AGNs with highly beamed relativistic jets pointing to the line of sight are, in fact, the most common sources of $\gamma-$rays.  Their strongly Doppler boosted non-thermal emission, with apparently very high fluxes, is power-law and usually attributed to relativistic particles 
accelerated stochastically in shocks  along the jet. However, the observed very short duration  of the flares (for instance, of  a few minutes only, in the case 
of  PKS2155-304), imply explosive and  extremely compact acceleration/emission regions ($< R_S /c$) with  Lorentz factors much larger than the typical bulk values
($\Gamma \sim 5-10$),  in order to avoid electron-positron pair creation  and therefore,  a  complete gamma-ray re-absorption within the source 
(e.g. \cite{Begelman08}). 
The only mechanism  able to explain this high variability   and compactness of the  TeV emission seems to be fast magnetic reconnection involving misaligned 
current sheets inside the jet \cite{giannios09,kushwaha17}. A similar mechanism has been also invoked to explain the transition from magnetically to kinetically
dominated flow and the prompt gamma-ray emission in gamma-ray-bursts (GRBs; \cite{zhangyan11}). Furthermore, the simultaneous detection of VHE gamma-ray flare 
and neutrino emission in TXS 0506+056 is suggestive of hadronic interactions between highly relativistic protons and background photons that result in pions
and a subsequent  decay in gamma-rays and neutrinos. If produced in the magnetically dominated regions of the blazar jet near the core, these protons are 
probably accelerated by magnetic reconnection \cite{degouveia18}.

%Another problem that is currently  challenging the researchers regards  the origin of the very high energy (VHE) emission of these sources. For instance,  until recently only AGNs with highly beamed jets pointing to the line of sight, namely  blazars,   were detected by gamma-ray telescopes.  More than a  chance coincidence, these detections  are consistent with the conventional scenario that attributes the VHE emission of these sources to particle acceleration along the jet being strongly Doppler boosted and producing  apparently very high fluxes. 

Also puzzling is the recent detection by  ground based gamma-ray observatories (e.g. \cite{sol13} and references therein) of TeV fluxes from non-blazar sources 
(i.e. with misaligned jets with the line of sight), belonging to the low luminosity AGN branch  for having  bolometric luminosities of only a few times the
Eddington luminosity (e.g. \cite{sol13} and references therein). These sources (usually callled LLAGNs) include well studied radio galaxies like  M$87$, 
Centaurus A, and Persus A.
%, and IC 310. 
Upper limit detections in the TeV range  from stellar mass galactic BHs of X-ray binary systems (also called BXBs), like Cyg X1 and X3,  have been also reported
(\cite{aleksic10,khiali15} and references therein). 
The angular resolution and  sensitivity of the current gamma-ray detectors are still  so  poor that it is hard to   establish  if this emission comes
from the jet or from the coronal region around the black hole (BH). Fortunately, the forthcoming ground based Cherenkov Telescope Array (CTA) 
(see \cite{actis11,acharya13}) will probably change this situation, but while it does not come, we may try at least to constrain the potential mechanisms. 
 %(e.g., \citealt{kachel10}). 
%These VHE detections were surprising because, besides being highly underluminous, the viewing angle of the jets of these sources is of several degrees, therefore allowing for only moderate Doppler boosting. These  characteristics make it hard to explain the VHE of these sources adopting  the same standard scenario of blazars. Furthermore, observations of  short time scale variability in the gamma-ray emission  of  IC 310, M87 and Per A   (e.g., \cite{ahar06,abdo09,ackermann12, aleksic14} and references therein) indicate that it  is produced in  a very compact region that could be perhaps the core.  In the case of Cen A, though  there is no evidence of significant variability at VHE, it has been also argued that   the TeV data of this source would be more compatible with a point source near the core  \cite{kachel10}. 
%If the gamma-ray photons were due, for instance,  to proton-proton  interactions along the jet then on leaving the source they would interact with the extragalactic background light (EBL) resulting in a flatter spectrum in the TeV range than the currently measured  with \textit{HESS} \cite{kachel09b}.

In this regard, an interesting evidence has been found by Kadowaki, de Gouveia Dal Pino \& Singh (2015) \cite{kadowaki15} who plotted in the same diagram the 
$\gamma-$ray luminosity versus the BH mass for about 230 sources including  non-blazar LLAGNs, black hole X-ray binaries (BXBs), blazars and GRBs, spanning  
$\sim 10$ orders of magnitude in mass and luminosity (see Figure 1).  This diagram evidences two distinct branches, or correlation trends. One branch is followed by blazars 
and GRBs, and the other by LLAGNs and BXBs. 
%(see Kadowaki et al. 2015, Singh, de Gouveia Dal Pino \& Kadowaki 2015). 
The association between blazars and GRBs is already expected as their non-thermal radiation is commonly attributed to Doppler boosted  emission by accelerated 
particles along the relativistic jets (e.g. \cite{nemmen12}) (as discussed above). But the lack of correlation  of these sources with  non-blazars and BXBs  in 
this diagram (see Figure 1) suggests that another location and/or population of relativistic particles may be producing the $\gamma-$ray emission in the latters.
In earlier work, de Gouveia Dal Pino \& Lazarian (2005) \cite{gl05} and de Gouveia Dal Pino et al. (2010) \cite{degouveia10a,degouveia10b}, inspired by the physical
processes in the solar corona,
had proposed a non-thermal emission model  based on the acceleration of particles by magnetic reconnection in the coronal region of accreting BH sources. 
This model, revisited in Kadowaki et al. (2015 \cite{kadowaki15}; see also \cite{singh15}), have demonstrated that the 
%There is direct evidence of this process  from GR-MHD simulations of accretion disks around BHs, as indicated by  Figure 2 (right).Fast magnetic reconnection between the lines of the BH magnetosphere and those arising from the accretion flow (with rates around 0.1 $V_A$; see Kadowaki, de Gouveia Dal Pino \& Stone in prep.) allows for the release of magnetic energy that heats and accelerates the plasma. 
 magnetic  reconnection power  released in fast events, driven by local turbulence embedded in  the large scale coronal magnetic lines, matches quite well with 
 the observed $\gamma-$ray luminosity  of the non-blazar LLAGNs and BXBs, plotted in the luminosity-BH mass diagram described above (see right panel of Figure 1, and  
 \cite{kadowaki15,singh15}). This result has been found to be nearly independent  of the detailed physics of the accretion flow and suggests that magnetic
 reconnection occurring in the core region of these sources can in principle explain the observed $\gamma-$ray emission.  
Though other studies have attempted to explain these observations as produced in the  jets of these sources (e.g. \cite{tavecchio08, lenain08}),  
the evidences above led to a more detailed investigation of particle acceleration scenarios involving the production of the VHE in the coronal region of 
the accreting  BH (see \cite{kadowaki15,khiali15,khiali16a} and  references therein). If such environments are magnetically dominated, as generally believed, 
then fast magnetic reconnection could be unavoidable, natural, and even efficient \cite{gl05, degouveia15, degouveia16, werneretal17}.

%Researchers have been facing  similar difficulties also  with the interpretation of the  VHE emission  of GBHs and their jets, particularly with the sources Cygnus X1 and X3 for which  recent detections indicate upper limits in the TeV range too  \cite{kadowaki15, singh15,khiali15}. 

%Even in the cases where the emission is most probably produced along the relativistic jets, as in blazars,  the conventional mechanism to accelerate the particles based on diffusive shock acceleration is facing severe constraints imposed by current VHE observations with very  high variability, of the order of  minutes in the TeV range. This implies extremely compact acceleration/ emission regions ($< R_S/c$) and besides,  it requires bulk Lorentz factors larger than the typical  values expected for such sources (around $\gamma \simeq 5-50$)  in order to avoid electron-positron pair creation. To circumvent these problems,  \cite{giannios09} proposed a reconnection model involving misaligned $mini-jets$ inside the jet. Fast reconnection has been also invoked in \cite{zhangyan11} to explain the transition from magnetically to kinetically dominated flow and the prompt gamma-ray emission in GRBs.

In this lecture, we discuss briefly particle acceleration by fast magnetic reconnection  and its potential role on the production of VHE in the surrounds of
BH sources and along their relativistic jets. We will also argue about neutrino emission counterparts.  Recent reviews on this subject can be also found in
 \cite{degouveia15,degouveia16, degouveia18, werneretal17} to mention a few. 
%accretion disk/jet systems around BHs  and show that this can be a powerful process to accelerate relativistic particles, produce the associated non-thermal emission and  particularly  the VHE one, and account for an efficient conversion of magnetic into kinetic energy.  

%FIGURE 1

\begin{figure}[h]
%    \begin{subfigure}[b]{0.5\textwidth}
    \includegraphics[width=3.0in]{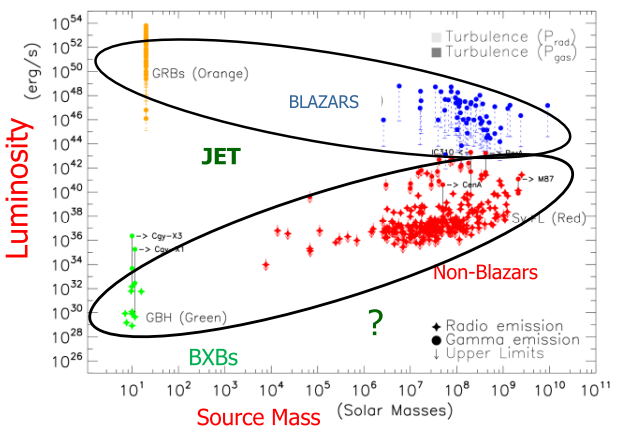}
%    \end{subfigure}
%    
%    \begin{subfigure}[b]{0.5\textwidth}
    \includegraphics[width=3.0in]{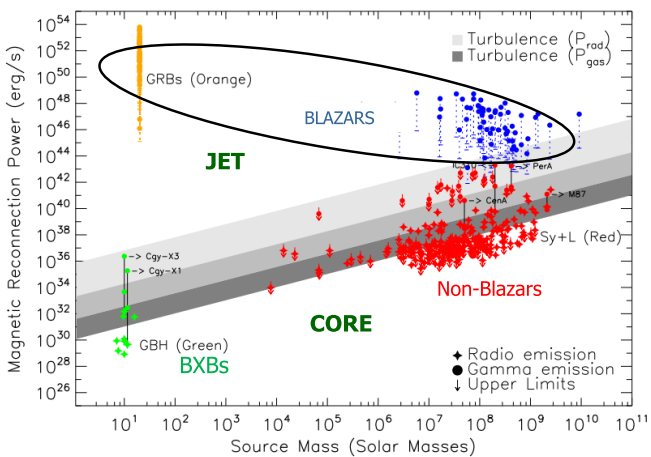}  
%    \end{subfigure}
\caption{
Observed radio and VHE emission of low luminous AGNs (LLAGNs, identified as non-blazars in the diagrams), galactic black hole binaries (BXBs, also labeled as GBHs in the diagrams), high luminous AGNs (blazars) and gamma ray burst (GRBs) versus BH source mass (symbols), compared to  the calculated turbulent driven magnetic reconnection power 
(gray region in the right diagram). 
The core radio emission of the BXBs (or GBHs) and non-blazars (or LLAGNs) is
represented by red and green diamonds, the gamma-ray emission of these two
classes is represented by red and green circles, respectively.  In the fewer sources
for which there is observed gamma-ray luminosity it is plotted the maximum and
minimum values linking both circles with a vertical black line that extends down
to the radio emission of each  of these sources. The inverted arrows associated
to some sources indicate upper limits in gamma-ray emission.  For blazars and
GRBs, only the gamma-ray emission is depicted, represented in blue and orange
circles, respectively. The vertical dashed lines correct the observed emission
by  Doppler boosting effects. The calculated magnetic reconnection power  clearly matches
the observed radio and gamma-ray emissions from LLAGNs (non-blazars) and GBHs (or GXBs), but not that
from blazars and GRBs. This result confirms early evidences that the emission
in blazars and GRBs is produced along the jet and not in the core of the sources, but on the other hand do indicate that the gamma-ray emission 
of LLAGNs and GBHs can be instead produced in the core region around the BH by magnetic reconnection processes (modified from \cite{kadowaki15, singh15}). }
%\label{figAGN-BHB}
\end{figure}

\section{Particle Acceleration by Fast Magnetic Reconnection}

There is fast magnetic reconnection  when two magnetic fluxes of opposite polarity encounter each other and partially annihilate at an efficient rate $V_R$ which
is a substantial fraction of the local Alfv\'en speed ($V_{A}$). Besides successful laboratory reconnection experiments \cite{yamada_etal_10} and direct 
observations in the earth magnetotail and in the solar corona flares, extensive  numerical work has been also carried out to understand the nature of this process,
both in collisionless (see, e.g. \cite{uzdensky15} for a review) and collisional plasmas (e.g., \cite{kowal09,loureiro07}). Different processes such as 
plasma instabilities, anomalous resistivity,
%(e.g., \cite{biskamp97,zenitani09}),  
and turbulence, can lead to fast reconnection. The latter process is  very efficient and probably the main driving mechanism of fast reconnection in collisional
MHD flows \cite{LV99,eyink11,takamoto15}.  
Turbulence causes the wandering of the magnetic field lines which allows for many independent patches to reconnect simultaneously making the reconnection rate 
large and independent on the local existing microscopic magnetic resistivity, $V_R \sim v_A (l_{inj}/L)^{1/2} (v_{turb}/v_A)^2$, where  $l_{inj}$ and  $v_{turb}$
are the injection scale and velocity of the turbulence, respectively \cite{LV99}.

The break of the magnetic field topoloy by fast reconnection involves the  release of  magnetic energy
explosively which explains, for instance, the bursty emission in  solar
flares, but it can also explain flares in compact sources like pulsar magnetospheres and winds, relativistic jets and BH coronae. Relativistic particles are 
always observed in connection with these
flares suggesting that magnetic reconnection can lead to direct particle
acceleration \cite{degouveia15}.

de Gouveia Dal Pino \& Lazarian \cite{gl05} seem to have been the first to realize that particles could be accelerated in magnetic reconnection sites in a similar
way as in shocks where particles confined
between the upstream  and downstream  flows undergo a first-order Fermi
acceleration. Likewise, trapped particles are also able to bounce
back and forth between the two converging magnetic fluxes   of a 
 reconnection discontinuity (or current sheet),  gyrorotate around the reconnected
magnetic field  (see Figure 2 in \cite{kowal11}, gaining energy due to
collisions with magnetic fluctuations at a rate $\Delta E/E \propto V_{R}/c$  {implying} a
first-order Fermi process with an exponential energy growth after several round
trips \cite{gl05}.  In Kowal et al. (2011) \cite{kowal11}, we have demonstrated the equivalence between this process and that of particles being accelerated while
confined within merging magnetic islands, as often seen in two-dimensional reconnection flows. 

The importance of reconnection acceleration as a key mechanism to explain observed non-thermal, highly variable emission, specially at VHEs from magnetically 
dominated sources, as described in Section 1,  has made this a field of intensive theoretical and  numerical study mainly through two-dimensional  particle-in-cell
(PIC) simulations of collisionless  plasmas  
\cite{drake06,zenitani01,zenitani07, zenitani08,lyubarsky08,drake10,clausen-brown2012, cerutti14, li15,  Lyutikovetal17, werneretal17, werneretal19}, 
and more recently also through three-dimensional (3D) PIC simulations \cite{sironi14, guo15, guo16}. However, these simulations, though essential in several 
aspects,  can probe acceleration only at the kinetic scales of the plasma, of a few hundreds of the inertial length   ($\sim 100 c/\omega_p$, where $\omega_p$ 
is the plasma frequency). To assess the  Fermi process in the large scales of the collisional MHD flows commonly observed in astrophyisical systems, the tracking
of test particle distributions in such flows is a very useful and complementary tool to help in the understanding of the overall process through the scales. 
Such studies   have been also successfully tested both in 2D and 3D MHD simulations with the injection of thousands of test particles in the reconnection domain 
by our and other groups \cite{kowal11, kowal12, delvalle16, onofri06,drake09,gordovskyy10,ding10}.  
 
Our studies, in particular,  have revealed that test protons injected in  large scale 3D MHD current sheets with turbulence embedded in order to make reconnection 
fast \cite{kowal12, delvalle16}, undergo  efficient particle acceleration 
with a rate $t_{acc}^{-1} \propto E^{-\alpha}$, with $0.2 < \alpha < 0.6$,  for a vast range of values of  $c / V_{\rm A} \sim 20 - 1000$.  We have obtained 
power-law spectra for the accelerated particles with power-law indicies  in the initial times of the simulations $N(E)\propto E^{-1,-2}$, where $E$ is the 
particle kinetic energy \cite{delvalle16}. These power-law indices are compatible with the values derived in PIC studies of highly relativistic magnetized,
non-radiative plasmas,  both of electron-positron pairs  (e.g, \cite{guo14,guo15,werneretal16}) and  electron-ion pairs 
(e.g., \cite{werneretal17,ball18a})\footnote{We remind that reconnection in electron-ion plasmas should be identical to that in electron-positron
plasmas in a highly relativistic regime, since the  energies become much larger than the rest mass energies both for electrons and ions in this case. In the
mild or semi-relativistic case, where electrons are highly relativistic and ions weakly relativistic, the PIC simulations of Werner et al. (2018)
\cite{werneretal17}, for instance, 
have revealed steeper power-law indices for the electron spectra, between 2 and 4, for decreasing plasma magnetization.}.  Our studies have also demonstrated the important
property that the reconnection acceleration rate is nearly independent of the intrinsic  parameters that drive the turbulence and thus the fast reconnection in the
collisional MHD flow. This is consistent with the notion that the turbulence is just the driving mechanism of fast reconnection and should not influence much the 
acceleration mechanism that is universal!  

The results summarized above are general and in principle applicable to a vast amount of magnetized astrophysical flows. In the following paragraphs, we will
discuss their  applications to relativistic jets and coronal accreting plasma around BH sources, highlighting interesting properties found in  our  recent 
studies. I also refer to the contributions of Kadowaki et al. (2019) \cite{kadowaki19} and Ramirez-Rodriguez et al. (2019) \cite{rodriguez19} in these Proceedings.  

\section{Magnetic Reconnection Acceleration and VHE in Relativistic Jets}

%As stressed in Section 1, fast magnetic reconnection may be an important mechanism to allow for the efficient release of magnetic energy and the transition from magnetically to kinetically dominated flow,  as well as allow for particle acceleration and the production of flaring  gamma-ray emission in blazars and GRBs. 

 The origin of relativistic jets near  central BHs is  still uncertain, but the usually  most accepted models rely on magnetic processes, like  magneto-centrifugal
 acceleration by helical magnetic fields arising from the accretion disk \cite{blandford_payne82}. Alternatively,  they can be powered by the BH spin transferred
 to the surrounding magnetic flow  (\cite{Blandford_Znajek1977}; see also Blandford 2019 in these Proceedings \cite{blandford18}). 
Thus in any case the  prediction is  that the jets should be  born as magnetically  or  Poynting flux dominated flows (see also \cite{degouveia18}).
The fact that at  observable distances from the source (starting  around $1000 R_S$ or less, where $R_S$ is the Schwartzschield radius) these jets become 
kinetically dominated, indicate that they should suffer efficient  conversion (or dissipation) of magnetic into kinetic energy. It has been argued by several 
authors, including ourselves (e.g., \cite{degouveia15,bromberg16})  that magnetic reconnection may be a powerful mechanism operating in these jets to allow for this conversion. 

To probe this process, we  performed  3D relativistic MHD simulations of rotating Poyinting flux dominated tower jets with initial helical fields \cite{singh16}.
Considering models with a ratio between the magnetic and the rest mass energy of the flow $\sigma  =1$,  and different density ratios between the jet and the 
environment, we induced precession perturbations that quickly developed current-driven kink (CDK) modes \cite{mizuno12,das18}. 
Figure 2 depicts a snapshot of a tower jet with  density larger than that of the environment, after the propagation of a helical kinked structure driven by the
CDK instability. This causes substantial distortion of the original straight plasma column leading to break and reconnection of the magnetic fields lines
in several regions.  The loci of maximum current density in this example are labeled with red circles. They trace the current-sheets where turbulent fast magnetic 
reconnection occurs with rates $\sim (0.01 - 0.15)V_A$ (\cite{singh16,kadowaki19}). We have employed the algorithm developed in 
Kadowaki et al. (2018) \cite{kadowaki18}  to localize these current sheets (see more details in  Kadowaki et al. 2019 in these Proceedings \cite{kadowaki19}). These values are compatible
with those of recent studies of fast turbulent relativistic reconnection in MHD current sheets \cite{takamoto15} and also in collisionless PIC simulations 
\cite{werneretal17,ball18b}. Moreover, these values produce similar  reconnection rates as in non-relativisitc flows \cite{kowal09,kadowaki18}.

Figure 3 shows the time evolution of the volume averaged magnetic and kinetic energy densities of the jet flow of Figure 2. We see that the CDK instability and the 
magnetic reconnection lead to release and  dissipation of magnetic energy accompanied by exponential increase of the kinetic energy density up to saturation 
(see also \cite{singh16}). We have also  found that this maximum growth  of the kinetic energy coincides with the maximum growth  of the rate of magnetic 
reconnection around the time interval $t= 40$ in code units (see Figure 3 of Kadowaki et al. 2019 in these Proceedings \cite{kadowaki19}).

As discussed in Section 2, these fast reconnection regions are potential  sites for Fermi particle acceleration. To probe this, we have performed $in$  $situ$ 
test particle acceleration injecting  1000 particles with initial Maxwellian velocity distribution into the relativistic jet of Figure 2 (see left diagram).  
Starting with energies $\sim 1$ MeV, the particles undergo an exponential acceleration when they are trapped in the reconnection sheets and interact resonantly
with plasma magnetic field fluctuations (as predicted in \cite{gl05}, and successfully tested in \cite{kowal11,kowal12,delvalle16}), until a saturation level
when then the accelerated particles Larmor radius become larger than the acceleration 
regions. The initial background magnetic field in this simulation had a maximum initial value $B=0.13$ G, but we have also considered simulations with values about
100 times larger. In the latter case, when  the particles enter the exponential regime, they  are accelerated up to energies $\sim 10^{19}$ eV at sub-pc
scales for an AGN jet (\cite{medina19}).
%(Medina-Torrejon, de Gouveia Dal Pino, Kowal, Mizuno, Singh, Kadowaki, in prep.).

The implications of these results are specially important for blazar jets, as argued in Section 1. These accelerated protons have energy enough to produce  VHE 
$\gamma-$rays within very compact regions as well as neutrinos, as recently detected  in the blazar TXS 0506+056 \cite{IceCube18}, a scenario that
deserves further detailed investigation in forthcoming work.

%FIGURE 2

\begin{figure}[h]
% \vspace*{-2.0 cm}
\begin{center}
 \includegraphics[width=5.5in]{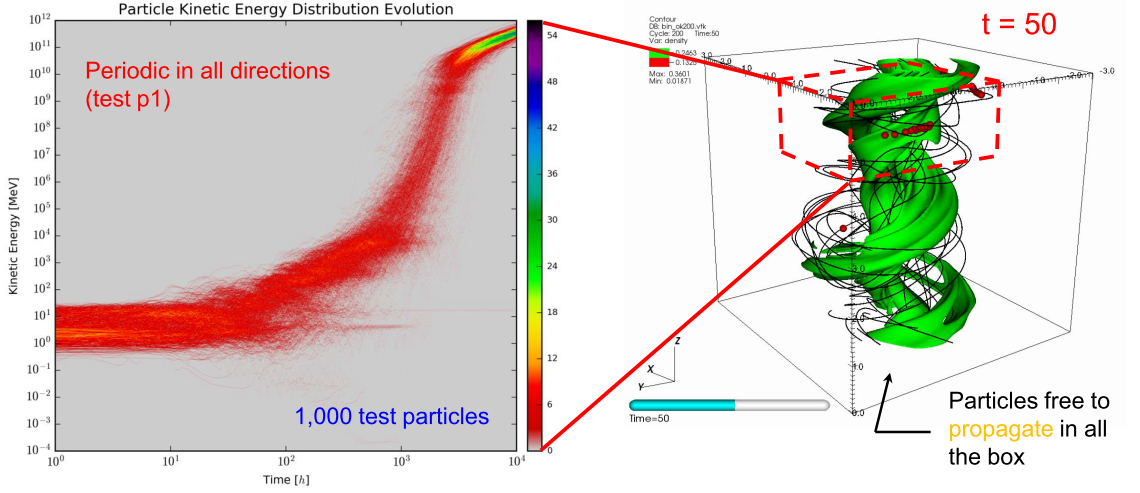} 
% \vspace*{-1.0 cm}
 \caption{The right diagram depicts 3D density isosurfaces (green) with superposed magnetic field lines (black) of a tower jet at time $t=50$ in code units ($L/c$)
 subject to the CDK instability that has distorted the originally straight plasma column with a helical magnetic field. The red circles identify the sites of fast reconnection
 driven by the CDK turbulence, employing the technique described in Kadowaki, de Gouveia Dal Pino \& Stone (2018 \cite{kadowaki18}; see also \cite{kadowaki19}). 
 In this simulation, we considered a rotating jet with angular velocity equal 2 in code units, and initial maximum magnetic field intensity $B= 0.13 G$. 
 The boundaries  are periodic in the z-direction, and outflow in the  x and  y directions. The grid resolution is 
 $L/40$, where L is the length scale code unit.  The velocity code unit is the light speed, and the magnetic field is in units of $(4\pi  \rho_o c^2)^{1/2}$, where $\rho_o$ is the ambient density,  taken as one code unit. 
The left diagram depicts  the time evolution of the kinetic energy of 1000 test particles (protons)  injected with an initial Maxwellian velocity distribution at 
the marked zone of the jet with a large concentration of current sheets (assuming periodic boundaries for the particles in all directions). Once particles are trapped within the reconnection regions they are exponentially accelerated 
up to energies $\sim 10^{17}$ eV in this model. Simulations performed with magnetic fields one-hundred  times larger, allow for particle acceleration up to $10^{19}$ eV
\cite{medina19}.}
\label{fig1}
\end{center}
\end{figure}

%FIGURE 3

\begin{figure}[!h]
\begin{center}
\includegraphics[angle=0,scale=0.5]{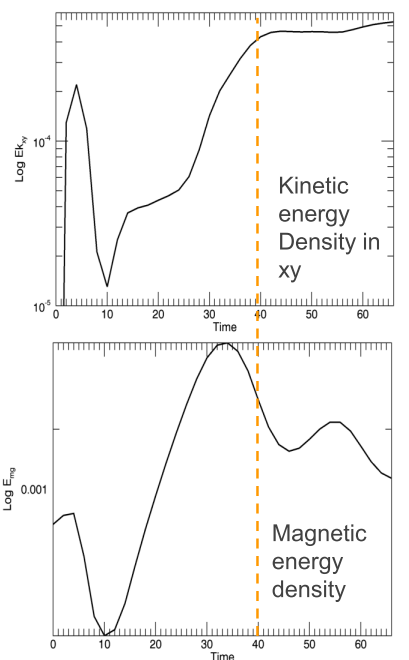}
\end{center}
\caption{Volume-averaged time evolution of the kinetic (top) and magnetic (bottom) energy densities for the
jet model presented in Figure 2 (\cite{medina19}; see also  \cite{singh16}).
}
% \label{f10_curlB_dec}
\end{figure}

\section{Magnetic reconnection acceleration and VHE around black holes}

As stressed in Section 1, fast magnetic reconnection can also occur in  magnetized accretion flows around the BH sources in both BXBs and  AGNs and this has 
been investigated by several authors (e.g., \cite{galeev79,dimatteo98,gl05,uzdensky08,degouveia10a,degouveia10b,hishino12,khiali15,kadowaki15,singh15}). 

We have obtained direct evidence of this process  from general relativistic MHD (GRMHD) simulations of torus accretion disks around BHs, using the ATHENA$++$ code 
\cite{white16}, as shown in Figure 2 of Kadowaki et al. (2019 in these Proceedings \cite{kadowaki19}; see also  \cite{parfrey15,dexter14,koide08,degouveia18,kadowaki18,ball18b}). Moreover, employing a 
similar algorithm for current sheet search as described in Section 3  (\cite{kadowaki18,kadowaki19}), we  could indentify several fast reconnection events in the 
turbulent accretion flow, with  rates  between 0.01 and 0.7 in  Alfv\'en velocity units. This is the first time this method is applied in GRMHD simulations 
(see more details in Kadowaki et al. 2019 in these Proceedings \cite{kadowaki19}), and it demonstrates the efficiency of the reconnection process, as well its potential to accelerate particles
and produce the VHE emission.
 
 Considering the approximate magnetic reconnection power that is produced  in the accreting flow around a BH, as calculated analytically in Kadowaki et al.  \cite{kadowaki15}
 and Singh et al.  \cite{singh15},  we computed the  
 spectral energy distribution (SED) of a number of  BXBs \cite{khiali15} and LLAGNs \cite{khiali16a,khiali16b} that lie  in the $"core"$ branch of Figure 1, aiming at explaining the origin of the VHE gamma-ray emission in these sources.
 Employing simple one-zone models to calculate the non-thermal leptonic and hadronic emission of the accelerated particles by reconnection, we could match the 
 observed spectrum and in particular, the VHE flux at TeV energies of the investigated sources (\cite{khiali15,khiali16a,khiali16b}). Nevertheless, these simple models, though elucidating in 
 many aspects,  did not account for the spatial distribution of the background fields nor provided an appropriate calculation of the gamma-ray absorption  by 
 electron-positron pair production.  
  
  We are currently developing more detailed models, combining three different techniques (see Figure 4 and also Ramirez-Rodriguez et al. 2019 in these Proceedings \cite{rodriguez19}). 
  First, we perform 2D GRMHD simulations of an accretion torus (using here the HARM code \cite{gammie03}) which allows us to obtain the density and magnetic field space distribution in the accretion flow around the BH (left column  in Figure 4). Then, we employ post-processing radiative transfer calculation (employing the GRMONTY code \cite{dolence09}) to obtain the background photon field due to leptonic (electronic) Synchrotron and Invervse Compton emissions (middle column in Figure 4). Having the space distributions of these three background fields, we inject a power-law distribution of relativistic 
  protons accelerated by magnetic reconnection according to the model described in Singh et al. (2015) \cite{singh15}. The interactions of these protons with the background
  fields, calculated using the CRpropa code (\cite{alves16}; see right column of Figure 4), allow us to obtain the space distributions and integrated fluxes of the hadronic emission, as well as the electron-positron pair creation and $\gamma$-ray absorption, and the neutrino emission. The application of this analysis to  Per A AGN \cite{rodriguez18},
  Cen A \cite{rodriguez19}, and to our galactic center Sgr A* \cite{rodriguez19b},  have demonstrated that the  TeV emission in these sources  can be explained by proton-proton and proton-photon interactions involving  accelerated protons by reconnection. More important, we have found that this emission is $not$ entirely absorbed by electron-positron pairs, as 
  suspected before in earlier studies of these sources.  Furthermore, the same process leads to neutrino emission too which can be compared directly 
  with the observed extragalactic diffuse emission by the IceCube (see \cite{khiali16b}). The details of this approach and the resulting 
  SED for CenA as an example,  are described in detail in Ramirez-Rodriguez et al. 2019, in these Proceedings \cite{rodriguez19}). 

%FIGURE 4

\begin{figure}[!h]
\begin{center}
\includegraphics[angle=0,scale=0.5]{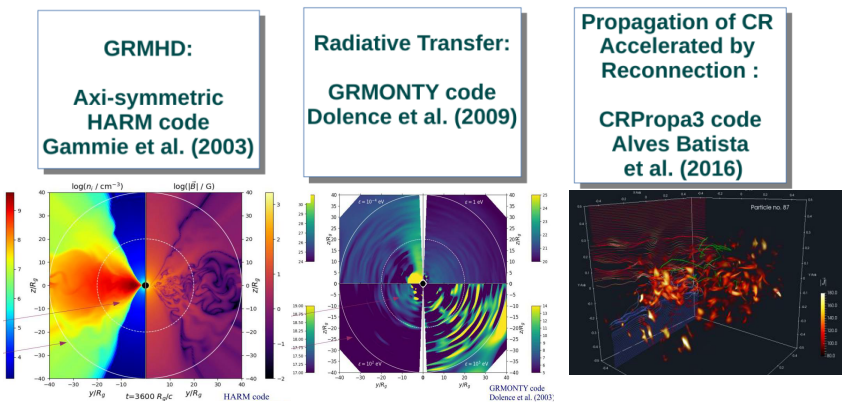}
\end{center}
\caption{Schematic representation illustrating the combination of three different techniques to calculate the spectral energy distribution (SED) around BH sources.
Left column depicts the density and magnetic field distribution of a 2D GRMHD simulation of a torus accreting flow around a BH where turbulence induced by
magneto-rotational instability drives magnetic reconnection (as shown in Kadowaki et al. 2019, these Proceedings \cite{kadowaki19}). Middle column depicts the leptonic radiation 
field resulting from a radiative transfer simulation of photons in the background of the accreting flow of the left diagram. Each quadrant shows 
a photon energy distribution, in the clockwise sense from the top-left: $10^{-1}$ eV, $1$ eV, $100$ eV, and $10^5$ eV \cite{rodriguez18}. 
Right column shows the trajectory of a single test particle (green line) being accelerated within a single current sheet by reconnected magnetic fields driven 
by turbulence \cite{delvalle16}. A similar process should happen with an entire distribution of particles injected in the accretion flow depicted in the 
left column that contains several regions like this. We then injected a distribution of accelerated protons by reconnection in the inner circle (described in the 
left diagram by a white line) and computed the spatial hadronic radiative losses distribution over the domain (contained within the outer white circle depicted in
the left diagram). See more details in \cite{rodriguez19,rodriguez19b}.}
% \label{figure4}
\end{figure}

\section{Summary and Conclusions}
In this lecture, we have discussed the importance of   turbulent magnetic reconnection and particle acceleration by reconnection  in  BH coronal regions and along relativistic jets based on studies supported by theory and numerical simulations of collisional MHD flows. Our results are consistent with those found in collisionless 
plasma studies of non-relativistic and relativistic reconnection, as stressed in Sections 1 and 2.
Our  main conclusions can be summarized as follows (see also \cite{degouveia18}):
%\item[Magnetic reconnection is important in accretion and jet systems for dissipation of magnetic energy, conversion into  kinetic energy, and particle acceleration.]	

\begin{itemize}
\item Fast magnetic reconnection driven by instabilities and turbulence can be an important process in accretion and jet systems to allow for an efficient dissipation of magnetic energy, and conversion into  kinetic energy and particle acceleration.

\item In magnetized plasmas, particles can be predominantly accelerated by fast magnetic reconnection (driven e.g.  by turbulence)   subject to a Fermi-like process,  and develop a power-law spectrum  $N(E) \sim  E^{-2,-1}$ \cite{kowal12,delvalle16}.  Steeper  power-law indices can result, particularly if radiative processes are present.

\item The magnetic reconnection power released in fast events occurring in the accretion flow around BHs \cite{kadowaki18,kadowaki19} could  explain the observed $\gamma-$ray  emission in non-blazar LLAGNs and BXBs.   This  power matches well with the observed correlation of $\gamma-$ray luminosity versus BH mass for these sources that spans 10 orders of magnitude \cite{kadowaki15,singh15}. We have also found that accelerated CRs by magnetic reconnection  in the core regions of these sources can produce TeV $\gamma-$rays via hadronic processes which  are not entirely re-absorbed by electron-positron pair production, as thought before. The same process also leads to neutrino production \cite{rodriguez18,rodriguez19,rodriguez19b}.
	  
\item In relativistic jets (as in blazars, GRBs and BXBs), magnetic reconnection acceleration may be also very efficient \cite{singh16,kadowaki19}. For instance, our numerical simulations of test particles injected in relativistic MHD tower jets have revealed the production of cosmic rays with energies up to $10^{17}$ eV and $10^{19}$ eV, for background magnetic fields around $\sim 0.13$ G and $\sim 13$ G, respectively \cite{medina19}.   These  protons have energy enough to be responsible for the $\gamma-$ray flares  and neutrino emission in the magnetically dominated regions of these jets. This  may be, for instance, the dominating mechanism responsible for the recent   observed $\gamma-$ray and neutrino emission in  the TXS 0506+056 blazar. This will be examined in detail in forthcoming work.

\end{itemize}
 
\section{Acknowledgments}
We acknowledge support from the Brazilian agencies FAPESP (2013/10559-5 grant) and CNPq (grant 308643/2017-8). The simulations presented in this lecture have made use of the computing facilities of the GAPAE group (IAG-USP) and the Laboratory of Astroinformatics IAG/USP, NAT/Unicsul (FAPESP grant 2009/54006-4).


\begin{thebibliography}{99}
%\bibitem{...}

\bibitem{actis11} Actis, M.  et al. (CTA collaboration) \ 2011, Exp. Astron., 32, 193

\bibitem{acharya13} Acharya, B.S. et al. (CTA collaboration) \ 2013,  Astropart. Phys., 43, 3

\bibitem{ahar07}Aharonian, F.~A., et al.\ 2007, The Astrophysical Journal, 664, L71

\bibitem{aleksic10} {Aleksi{\'c}}, J., {Antonelli}, L.~A., {Antoranz}, P., et al.\ 2010, ApJ, 721, 843

\bibitem{alves16} {{Alves Batista}, R., {Dundovic}, A., {Erdmann}, M., 
	{Kampert}, K.-H., {Kuempel}, D., {M{\"u}ller}, G., 
	{Sigl}, G., {van Vliet}, A., {Walz}, D., \& {Winchen}, T.
	},\ 2016, Jcap, 5, 038 

\bibitem{ball18a} {{Ball}, D., {Sironi}, L. \& {{\"O}zel}, F.}\ 2018a, ApJ, 862, 80

\bibitem{ball18b}  {{Ball}, D., {{\"O}zel}, F., {Psaltis}, D., {Chan}, C.-K. \&
	{Sironi}, L.}\ 2018b, ApJ, 853, 184

\bibitem{Begelman08} {Begelman}, M.~C., {Fabian}, A.~C. \& {Rees}, M.~J.\ 2008, MNRAS, 384, L19

\bibitem{blandford18} {{Blandford}, R. D., {Meier}, D. \& {Readhead}, A.}\ 2018, arXiv:1812.06025

\bibitem{blandford_payne82} Blandford, R.~D., \& Payne, D.~G.\ 1982, MNRAS, 199, 883

\bibitem{Blandford_Znajek1977} Blandford, R. D., \& Znajek, R. L.\ 1977, MNRAS, 179, 433

\bibitem{bromberg16} Bromberg, O. \& Tchekhovskoy, A.\ 2016, MNRAS, 456, 1739 

\bibitem{cerutti14} Cerutti, B., Uzdensky, D.~A. \& Begelman, M.~C. \ 2014, ApJ, 746, 148

\bibitem{clausen-brown2012} Clausen-Brown, E. \& Lyutikov, M. \ 2012, MNRAS, 426, 1374

\bibitem{das18} Das, U., \& Begelman, M. C.\ 2018, arXiv:1807.11480
 
\bibitem{degouveia15} de Gouveia Dal Pino, E.~M., \& Kowal, G. 2015, in Magnetic Fields in Diffuse Media, Astrophysics and Space Science Library, Eds. A. Lazarian. E. de Gouveia Dal Pino, C. Melioli,  407, 373

\bibitem{degouveia18}{{de Gouveia Dal Pino}, E. M., {Kowal}, G., {Kadowaki}, L., {Medina Torrej{\'o}n}, T.~E., {Mizuno}, Y. \& {Singh}, C.},\ 2018, arXiv:1809.06742v1

\bibitem{gl05} de Gouveia Dal Pino, E. M., \& Lazarian, A. 2005, A\&A, 441, 845 (GL05)

\bibitem{degouveia10a} de Gouveia Dal Pino, E.~M., Piovezan, P.~P. \& Kadowaki, L.~H.~S.\ 2010a, A\&A, 518, A5

\bibitem{degouveia10b} de Gouveia Dal Pino, E.~M., Piovezan, P.~P., Kadowaki, L.~H.~S., Kowal, G. \&  Lazarian, A. \ 2010b, Highlights of Astronomy, 15, 247

\bibitem{degouveia16} {de Gouveia Dal Pino}, E., {del Valle}, M.~V., {Kadowaki}, L., {Khiali}, B., {Kowal}, G., {Mizuno}, Y. \& {Singh}, C.~B. \ 2016, arXiv:1608.03173

\bibitem{delvalle16} del Valle, M. V., de Gouveia Dal Pino, E.M., \& Kowal, G. 2016, MNRAS, 463, 4331

\bibitem{dexter14} Dexter, J., McKinney, J.~C., Markoff, S., \& Tchekhovskoy, A.\ 2014, MNRAS, 440, 218

\bibitem{dimatteo98} {{Di Matteo}, T.}\ 1998, MNRAS, 299, L15

\bibitem{ding10} Ding J., Yuan F.,  \& Liang E., \ 2010, ApJ, 708, 1545

\bibitem{dolence09} {{Dolence}, J.~C., {Gammie}, C.~F., {Mo{\'s}cibrodzka}, M. \&
	{Leung}, P.~K.}\ 2009, ApJs, 184, 387

\bibitem{drake06} Drake, J.~F., Swisdak, M., Che, H., \& Shay, M.~A.\ 2006, Nature, 443, 553

\bibitem{drake10} Drake, J.~F., Opher, M., Swisdak, M, \& Chamoun, J.~N. \ 2010, ApJ, 709, 963

\bibitem{drake09} {{Drake}, J.~F., {Cassak}, P.~A., Shay, M.~A., {Swisdak}, M. \& {Quataert}, E.}\ 2009, ApJL, 700, L16

\bibitem{eyink11} {Eyink}, G.~L., {Lazarian}, A. \& {Vishniac}, E.~T.\ 2011, ApJ, 743, 51 

\bibitem{galeev79} Galeev, A. A., Rosner, R. \& Vaiana, G. ~S. \ 1979, ApJ, 229, 318

\bibitem{gammie03} {{Gammie}, C.~F., {McKinney}, J.~C. \& {T{\'o}th}, G.}\ 2003, ApJ, 589, 444 

\bibitem{giannios09} Giannios, D., Uzdensky, D. A. \& Begelman, M. C., \ 2009, MNRAS,  228, 395 L29-L33

\bibitem{gordovskyy10}  {{Gordovskyy}, M., {Browning}, P.~K. \& {Vekstein}, G.~E.
}\ 2010, ApJ, 720, 1603

\bibitem{guo14} {{Guo}, F., {Li}, H., {Daughton}, W. \& {Liu}, Y. H.},\ 2014, Physical Review Letters, 113, 155005

\bibitem{guo15} Guo, F., Liu, Y.~H., Daughton, W. \& Li, H.\ 2015, ApJ, 806, 167

\bibitem{guo16}  {{Guo}, F., {Li}, H., {Daughton}, W., {Li}, X. \& {Liu}, Y.-H.
	}, Physics of Plasmas, 23, 055708 

\bibitem{hishino12} {{Hoshino}, M. \& {Lyubarsky}, Y.}\ 2012, Space Science Reviews, 173, 521 

\bibitem{IceCube18} IceCube Collaboration\ 2018, Science, 361, 147 

\bibitem{kadowaki15} Kadowaki, L.~H.~S., de Gouveia Dal Pino, E.~M., \& Singh, C.~B.\ 2015, ApJ, 802, 113 

\bibitem{kadowaki18} {{Kadowaki}, L.~H.~S., {de Gouveia Dal Pino}, E.~M. \& {Stone}, J.~M.
	}\ 2018, ApJ, 864, 52
	
\bibitem{kadowaki19} {Kadowaki}, L.~H.~S., de Gouveia Dal Pino, E.~M., \& Medina-Torrejon, T.~E. \ 2019, these Proceedings

\bibitem{khiali16a} Khiali, B., de Gouveia Dal Pino, E. M. \& Sol, H.\ 2016, A\&A, arXiv:1504.07592

\bibitem{khiali16b} Khiali, B. \& de Gouveia Dal Pino, E.~M. \ 2016, MNRAS, 455, 838

\bibitem{khiali15} Khiali, B., de Gouveia Dal Pino, E. M., \& del Valle, M.~V.\ 2015, MNRAS, 449, 34

\bibitem{koide08} {{Koide}, S. \& {Arai}, K.}\ 2008, ApJ, 682, 1124

\bibitem{kowal09} Kowal, G., Lazarian, A., Vishniac, E.~T., \& Otmianowska-Mazur, K., 2009, ApJ, 700, 63
 
\bibitem{kowal11} Kowal, G., de Gouveia Dal Pino, E. M., \& Lazarian, A. 2011, ApJ, 735, 102

\bibitem{kowal12} Kowal, G., de Gouveia Dal Pino, E. M., \& Lazarian, A. 2012, PRL, 108, 241102

\bibitem{kushwaha17} Kushwaha, P., Sinha, A., Misra, R., Singh, K.~P.,  de Gouveia Dal Pino, E.~M. et al. \ 2017, ApJ, 849, 138

\bibitem{LV99}Lazarian, A. \& Vishniac, E.~T.\ 1999, ApJ, 517, 700

\bibitem{lenain08} Lenain, J.~P., Boisson, C., Sol, H. \& Katarzynski, K.\ 2008 A\&A, 478, 111

\bibitem{li15} Li, X., Guo, F., Li, H., \& Li, G.\ 2015, ApJ Lett., 811, L24

\bibitem{loureiro07} Loureiro, N.~ F., Schekochihin, A.~A., \& Cowley, S. C.\ 2007, Physics of Plasmas, 14, 100703

\bibitem{Lyutikovetal17}{{Lyutikov}, M., {Sironi}, L., {Komissarov}, S.~S. \&
	{Porth}, O.}, 2017, Journal of Plasma Physics, 83, 635830602

\bibitem{lyubarsky08} Lyubarsky, Y. \&  Liverts, M.\ 2008, ApJ, 682, 1436

\bibitem{medina19} Medina-Torrejon, T.~E.,  de Gouveia Dal Pino,  E.~M., Kowal, G.,  Mizuno, Y., Singh, C., \&  Kadowaki, L. \ 2019 (in prep.)

\bibitem{mizuno12} Mizuno, Y., Lyubarsky, Y., Nishikawa, K.-I., \& Hardee, P. E. \ 2012, ApJ, 757, 16

\bibitem{nemmen12} {{Nemmen}, R.~S., {Georganopoulos}, M., {Guiriec}, S., {Meyer}, E.~T., {Gehrels}, N. \& {Sambruna}, R.~M.}\ 2012, Science, 338, 1445

\bibitem{onofri06}  {{Onofri}, M., {Isliker}, H. \& {Vlahos}, L.}\ 2006, Physical Review Letters, 96, 151102

\bibitem{parfrey15} {{Parfrey}, K., {Giannios}, D., {Beloborodov}, A.~M.}\ 2015, MNRAS, 446, L61

\bibitem{rodriguez18} Rodriguez-Ramirez, J. C., de Gouveia  Dal Pino, E. M., \& Alves Batista, R.\ 2018, arXiv:1811.02812
 
\bibitem{rodriguez19} Rodriguez-Ramirez, J. C., de Gouveia Dal Pino, E. M., \& Alves Batista, R.\ 2019, these Proceedings

\bibitem{rodriguez19b} Rodriguez-Ramirez, J. C., de Gouveia Dal Pino, E. M., \& Alves Batista, R.\ 2019 (submitted)

\bibitem{singh15} Singh, C.~B., de Gouveia Dal Pino, E.~M., \& Kadowaki, L.~H.~S.\ 2015, ApJ Letts., 799, L20

\bibitem{singh16} Singh, C.~B., Mizuno, Y., \& de Gouveia Dal Pino, E.~M.\ 2016, ApJ, 824, 48 

\bibitem{sironi14} Sironi, L., \& Spitkovsky, A.\ 2014, ApJ Letts., 783, L21 

\bibitem{sol13} Sol, H., Zech, A., Boisson, C. et al.\  2013, Astroparticle Physics, 43, 215

\bibitem{tavecchio08} Tavecchio F., Ghisellini G.,\ 2008, MNRAS, 385, L98

\bibitem{takamoto15} Takamoto, M., Inoue, T., \& Lazarian, A.\ 2015, ApJ, 815, 16

\bibitem{uzdensky15} Uzdensky, D. A.\ 2015, arXiv:1510.05397

\bibitem{uzdensky08} {{Uzdensky}, D.~A. \& {Goodman}, J.}\ 2008, ApJ, 682, 608

\bibitem{werneretal16} {{Werner}, G.~R., {Uzdensky}, D.~A., {Cerutti}, B., 
	{Nalewajko}, K. \& {Begelman}, M.~C.}\ 2016, ApJL, 816, L8

\bibitem{werneretal17}  Werner, G.~R., Uzdensky, D.~A., Begelman, M.~C., Cerutti, B., \& Nalewajko, K.\ 2018, MNRAS, 473, 4840 

\bibitem{werneretal19} {{Werner}, G.~R., {Philippov}, A.~A. \& {Uzdensky}, D.~A.
	},\ 2019, MNRAS, 482, L60

\bibitem{white16} {{White}, C.~J., {Stone}, J.~M. \& {Gammie}, C.~F.}\ 2016, ApJs, 225, 22

\bibitem{yamada_etal_10}Yamada, M., Kulsrud, R., \& Ji, H.\ 2010, Reviews of Modern Physics, 82, 603

\bibitem{zenitani01}Zenitani, S. \& Hoshino, M.\ 2001, ApJ, 562L, 63Z

\bibitem{zenitani07}Zenitani, S. \& Hoshino, M.\  2007, ApJ, 670, 702

\bibitem{zenitani08}Zenitani, S. \& Hoshino, M. \ 2008, AJ, 677, 530

\bibitem{zhangyan11} Zhang, B., \& Yan, H. 2011, ApJ, 726, 90

%\bibitem{uzdensky11}

\end{thebibliography}
\end{document}